\documentclass[twocolumn,superscriptaddress,showpacs,pra]{revtex4}

\usepackage[utf8x]{inputenc}

\usepackage{epsfig}
\usepackage{epstopdf}
\usepackage{color}
\usepackage{delarray}
\usepackage{amsmath, amssymb}
\usepackage{bm}
\usepackage{float}
\usepackage{graphicx}
\usepackage{color}

\begin{document}
\title{Zeno Dynamics of Quantum Chirps}

\author{Cihan Bay\i nd\i r}
\email{cbayindir@itu.edu.tr}
\affiliation{Associate Professor, Engineering Faculty,  \.{I}stanbul Technical University, 34469 Maslak, \.{I}stanbul, Turkey \\
						 Adjunct Professor, Engineering Faculty, Bo\u{g}azi\c{c}i University, 34342 Bebek, \.{I}stanbul, Turkey \\
						 International Collaboration Board Member, CERN, CH-1211 Geneva 23, Switzerland}

\begin{abstract}
We investigate the Zeno dynamics of quantum chirps. More specifically, we analyze the Zeno dynamics of a chirped solitary wave solution of the non-dimensional nonlinear Schrödinger equation (NLSE) with a gain/loss term. We show that the motion of the quantum chirp can be inhibited by frequent measurements in the observation domain. In order to assess the effect of Zeno dynamics on the quantum chirp imaging resolution, we analyze and compare its point spread functions (PSF) under free evolution and under Zeno dynamics. Additionally, we investigate the effect of observation frequency on the quantum chirp profile as well as on its probability of lingering in the observation domain. Our results can be immediately used to investigate the signal properties of quantum chirps and their possible jamming properties by Zeno dynamics in quantum radar technology. Additionally, many potential applications in atomic physics and optics including nonlinear phenomena and chirps can be investigated using the approach presented in our paper.


\end{abstract}
\maketitle

\section{Introduction}
One of the most challenging problems in nonlinear quantum mechanics and nonlinear optics is the generation of short pulses. Short pulses are required for better signal and image resolution, however, in order to satisfy a certain average power limit these pulses can not be infinitely short. Researchers have overcome this problem using alternative approaches such as using linear frequency modulated codes, Barker codes, chirp-like phase codes, Golomb's codes, Huffman codes and asymptotically perfect codes \cite{Carrara, Collins, Curlander, Hein, Levanon, Soumekh1996, Soumekh2004, bay2013phd}. While some of these techniques are widely adopted and used in classical radar imaging, researchers followed different paths in optics. Pulse compression by means of higher-order soliton compression or adiabatic pulse compression were two of the widely utilized techniques in optics \cite{Moores96, QianL2009}. Although the higher-order soliton compression technique can achieve the large degrees of compression, it causes significant pedestal generation \cite{QianL2009}. Whereas in adiabatic compression, the pulse having a waveform of a soliton is used and a dispersion variation along the propagation direction is introduced, which allows the soliton to self-adjust by balancing dispersion and nonlinearity \cite{QianL2009}. Although this compression scheme is more successful in maintaining the transform-limited characteristics of the pulse, it is slower \cite{Moores96, QianL2009}. Moores showed that exact chirped soliton solutions of the nonlinear Schrödinger equation (NLSE) with a gain/loss exist, which allowed the rapid compression of the pulses as a remedy to this drawback \cite{Moores96}. Thereafter, some other analytical chirped soliton solutions of the NLSE and its extensions are reported in the literature \cite{QianL2009, Kruglov2004, Rosenthal2006}. Additionally, some other pulse types such as the self-similar pulses are also introduced and studied in the relevant literature for effectively generating short pulses \cite{Moores96, QianL2009}.

On the other hand, there is a vast amount of literature about the quantum Zeno dynamics, which can simply be defined as the `inhibition of the evolution of an unstable quantum state by appropriate frequent observations' \cite{MisraSudarshan,Facchi2008JPA}. Quantum Zeno effect and dynamics states that observations alter or can even stop the evolution of an atomic particle \cite{HarochePRL2010, HarochePRA2012, HarocheNP2014}. This effect is usually employed to freeze the decay of the quantum state of atomic particles. In this context, it is shown by various researchers that the Zeno effect can be used to prevent the dissipation of particles \cite{Guo1998PRL,Guo1998PRA}, to suppress the decoherence of qubit systems and entanglement of atoms \cite{Lloyd1998PRA, Maniscalco2008PRL} and to freeze the coherent field growth in cavities \cite{Bernu2008PRL}. It is also shown that the quantum Zeno effect and dynamics can also be used for entanglement creation, controlling operations and purification of the quantum systems \cite{Chen2016OptComm, Barontini2015Science, Nakazoto20013PRL}. Additionally, the linear and nonlinear optical analogues of the quantum Zeno effect became increasingly more popular \cite{Abdullaev2011PRA}. In this context, Zeno effect in optical fibers are reported in \cite{Yamane2001OptComm}. Controlling the optical gates \cite{Leung2006PRA}, optical switching \cite{Kumar2013PRL} and the nonlinear optical couplers \cite{Thapliyal2016PRA} by Zeno effect are just few of the problems tackled in these analogue optical systems. The usage of the Zeno effect in STM is also investigated \cite{Biagioni2008OptExp}.

To our best knowledge, the Zeno dynamics of quantum chirps and their optical analogs are not studied in the literature. With this motivation, we study the Zeno dynamics of quantum chirps and their optical analogs in this paper. More specifically, we analyze the Zeno dynamics and lingering probabilities of the quantum chirp soliton solution of the NLSE first introduced by Moores in \cite{Moores96}. We implement a spectral scheme with a $4^{th}$ order Runge-Kutta time integrator for the numerical solutions of the NLSE, and numerically investigate the Zeno dynamics of this quantum chirp using this scheme. We discuss the point spread functions (PSFs), lingering properties and possibilities of quantum chirps under Zeno dynamics and compare our results with some of the analytical results existing in the literature. We comment on our findings and discuss the limitation and possible usage of our results and method.

\section{Quantum Chirps of the Nonlinear Schr\"{o}dinger Equation and Their Zeno Dynamics}
Besides nonlinear optics, nonlinear quantum mechanics is becoming increasingly more important and popular. Many important and fascinating phenomena studied in these and similar branches of nonlinear science are studied within the frame of nonlinear Schr\"{o}dinger equation (NLSE) or its extensions which can adequately describe all the features of linear quantum mechanics \cite{Richardson2014PRA}. These phenomena include but are not limited to quantum optical solitons in photonic waveguides with various forms of nonlinearity \cite{Carter1987PRL,Drummond1987JOSAB,Lai1989PRA844, Bayrinp}, optical and quantum rogue waves \cite{Akhmediev2009b, Bay2020physca, BayPRE1, BayPRE2, BayOz}, nonlinear quantum entanglement \cite{Wang_Nl_entang}, noise induced amplification and tunneling \cite{Chia_Vedral, Bay_TWMS2017}, just to name a few. Various chirped solutions of the NLSE or its extensions are also reported in the literature \cite{Moores96, QianL2009, Kruglov2004, Rosenthal2006}. These chirped solutions of the NLSE promotes the possibility of quantum signal and image processing studies and technology, including quantum radar imaging.

In this paper, our motivation is to study the dynamics of quantum chirps when they evolve freely and when they are under the effect of Zeno dynamics. With this motivation, for modeling the Zeno dynamics of quantum chirps in nonlinear optics and quantum mechanics, we consider the non-dimensional NLSE with a gain/loss term given as
\begin{equation}
i\psi_t + \frac{1}{2} \psi_{xx} +  \left|\psi \right|^2 \psi-i\frac{\alpha(t)}{2} \psi =0,
\label{eq01}
\end{equation}
where $\psi$ denotes the complex amplitude and $i$ denotes the imaginary unity. In here, $x$ and $t$ are the spatial and temporal variables, respectively. In nonlinear optics, these variables are often switched with $t$ and $z$ variables, respectively, where $z$ denotes the propagation direction along the optical media. It is very well known that NLSE given by Eq.(\ref{eq01}) admits many different types of exact solutions including but are not limited to solitary waves, rational solutions, kinks, rogue waves and chirps. Some of the quantum chirp solutions of the NLSE are reviewed and analyzed in the coming sections of this paper. Although the analytical forms of the quantum chirps are known, under the effect of Zeno dynamics chirped wavefunction may become arbitrary, thus numerical solution of the NLSE is needed. With this motivation, we implement a spectral scheme with a $4^{th}$ order Runge-Kutta time integrator and solve the NLSE given in Eq.(\ref{eq01}) using this scheme. In order to perform the time stepping using  $4^{th}$ order Runge-Kutta algorithm, one can rewrite the NLSE given in Eq.(\ref{eq01}) as
\begin{equation}
\psi_t =\frac{i}{2} \psi_{xx} + i \left|\psi \right|^2 \psi + \frac{\gamma(t)}{2} \psi =h(\psi, t,x)
\label{eq02}
\end{equation}
where the RHS is denoted by a new function, $h(\psi, t,x)$ \cite{bayindir2016, demiray, Karjadi2012, trefethen, bayindir2016nature}. The second order spatial derivative in this formula can be calculated spectrally using the Fourier series at each time step using
\begin{equation}
\psi_{xx} = F^{-1} \left[ -k^2 F[\psi] \right]
\label{eq03}
\end{equation}
In here $F$ and $F^{-1}$ denote the Fourier and the inverse Fourier transform operations, respectively \cite{canuto, trefethen, Bracewell}. In Eq.(\ref{eq03}), $k$ denotes the wavenumber vector which includes exact $M$ multiples of the fundamental wavenumber, $k_o=2 \pi/L_d$. The domain for the computations is selected to be $x \in [-60,60]$, thus the domain length becomes $L_d=120$.  In order the perform FFTs efficiently, $M=1024$ spectral components are used for all the simulations presented in this paper. The nonlinear products are obtained by simple multiplication operation in the physical domain. For stable time stepping purposes, the four slopes of the $4^{th}$ order Runge-Kutta scheme can be calculated by
\begin{equation}
\begin{split}
& s_1=h(\psi^j, t^j, x) \\
& s_2=h(\psi^j+0.5 s_1dt, t^j+0.5dt, x) \\
& s_3=h(\psi^j+0.5 s_2dt, t^j+0.5dt, x) \\
& s_4=h(\psi^j+s_1dt, t^j+dt, x) \\
\end{split}
\label{eq04}
\end{equation}
In here, $dt$ denotes the time step which is selected as $dt=10^{-5}$ to avoid stability problems. Then, the time stepping of the complex wavefunction, $\psi$, and time parameter, $t$, are performed using the equations
\begin{equation}
\begin{split}
& \psi^{j+1}=\psi^{j}+(s_1+2s_2+2s_3+s_4)/6 \\
& t^{j+1}=t^j+dt\\
\end{split}
\label{eq05}
\end{equation}
iteratively, where $j$ denotes the iteration count. Therefore, using the numerical scheme summarized above, it is possible to model the temporal evolution of the complicated solutions of the NLSE given by Eq.(\ref{eq01}). 

In order to investigate the transient Zeno dynamics of quantum chirps, it is possible to use some analytical chirp solutions of Eq.(\ref{eq01}) existing in the literature, such as those listed in \cite{Moores96, QianL2009, Kruglov2004, Rosenthal2006}. In this study, the initial condition for the numerical scheme summarized above is selected to be the analytical chirp solution of the NLSE  given in \cite{Moores96} which exactly satisfies Eq.(\ref{eq01}). Following \cite{Moores96}, the analytical form of this quantum chirp can be written as
\begin{equation}
\psi(t,x) = \frac{A\alpha}{\alpha_0} \exp{\left[ i\phi(t,x) \right]} \textnormal{sech}{\left( \frac{A\alpha}{\alpha_0}x \right)},
\label{eq06}
\end{equation}
where 
\begin{equation}
\alpha(t) = \frac{1}{\alpha_0^{-1}-t}
\label{eq07}
\end{equation}
and 
\begin{equation}
\phi(t,x) =\phi_0- \frac{\alpha}{2} \left[x^2- \left(\frac{A}{\alpha_0}\right)^2 \right] 
\label{eq08}
\end{equation}
This quantum chirp is a frequency modulated waveform with a $sech$ type envelope. Again, following \cite{Moores96}, the value of $\alpha_0=9.5$ is used throughout this study. The value of $\phi_0=0$ is used in our simulations which ensures the symmetry of the chirp about the origin of the spatial domain. Using the quantum chirp specified by Eqs.(\ref{eq06})-(\ref{eq08}) as the initial condition, the time stepping is performed for two cases; the free evolution case and evolution under Zeno observations case.

It is well-known that the decay of an atomic particle can be inhibited by Zeno dynamics \cite{HarochePRL2010, HarochePRA2012, HarocheNP2014}. However, it remains an open question whether the quantum chirps or the chirped optical waveforms in the quantized optical fields can be stopped using Zeno dynamics. In this paper, we aim to address this open problem. Since the chirped waveform becomes complicated after a positive Zeno measurement, we use the numerical solution scheme of the NLSE summarized above. In order to model the effect of Zeno dynamics on the chirped waveform, we use a theoretical formulation first proposed in \cite{PorrasFreeZeno}. This formulation can be summarized as follows: after a positive Zeno measurement, the particle lingers in the observation domain of $[-L,L]$ for which the wavefunction becomes $\psi_T(x,t)=\psi(x,t)\textnormal{rect}(x/L)/\sqrt{P}$ where $P=\int_{-L}^L \left| \psi \left(x,t \right) \right|^2 dx$, and $\textnormal{rect}(x/L)=1$ for $-L \leq x \leq L$, and $0$ outside this observation domain \cite{PorrasFreeZeno}. The waveform under investigation evolves according to NLSE given by Eq.(\ref{eq01}), between two successive positive Zeno measurements. The summary of this cycle can be given as
\begin{equation}
\begin{split}
\psi_T &  \left(x,\frac{(n-1)t}{N} \right) \stackrel{evolve}{\rightarrow} \psi \left(x,\frac{nt}{N} \right) \stackrel{measure}{\rightarrow} \\
&    \psi_T \left(x,\frac{nt}{N} \right)= \psi \left(x,\frac{nt}{N} \right) \frac{\textnormal{rect(x/L)}}{\sqrt{P_N^{n}}}
\label{eq09}
\end{split}
\end{equation}
In here, $n$ denotes the observation index and $N$ denotes the number of observations \cite{PorrasFreeZeno}. The probability of lingering of the waveform in the observation domain after $n$ successive positive measurements becomes
 \begin{equation}
P_N^{n}=\int_{-L}^L \left| \psi \left(x,\frac{nt}{N} \right) \right|^2 dx.
\label{eq10}
\end{equation}
Using this result, the cumulative probability of lingering of the waveform in the observation domain $[-L,L]$ can be calculated as 
 \begin{equation}
P_N=\prod_{n=1}^N P_N^{n}.
\label{eq11}
\end{equation}
as discussed in \cite{PorrasFreeZeno}. Additionally, some analytical estimations for the probabilities given by Eqs.(\ref{eq10}) and (\ref{eq11}) are proposed in \cite{PorrasFreeZeno}. Utilizing the momentum representation of linear quantum mechanics and the analogy of optical waves of Fabry-Perot resonator, the analytical form for the lingering probability of an atomic particle in the observation interval of $[-L,L]$ after $n^{th}$ measurement is given as
 \begin{equation}
P_N^{n}\approx 1-0.12 \left(\frac{4}{\pi} \right)^2  \left(\frac{2 \pi t}{N} \right)^{3/2}
\label{eq12}
\end{equation}
in \cite{PorrasFreeZeno}. Similarly, the cumulative probability of lingering of the particle in the observation domain after $N$ measurements becomes 
 \begin{equation}
P_N \approx \left(1-0.12 \left(\frac{4}{\pi} \right)^2  \left(\frac{2 \pi t}{N} \right)^{3/2} \right)^N
\label{eq13}
\end{equation}
as discussed in \cite{PorrasFreeZeno}. It is possible to simplify these expressions further using Newton's binomial theorem \cite{PorrasFreeZeno}. This approach, which is proposed to model the effect of Zeno observations on the wavefunction, is also used in some other studies \cite{Porras_Zenotunnelmimick, Porras_diffractionspread, BayOz} and experimentally tested in \cite{Porras_zeno_clss_opt}. In this paper, we follow the same approach to model and analyze the Zeno dynamics and the lingering probabilities of the quantum chirps in the observation domain.

\section{Results and Discussion}

\begin{figure*}[ht!]
\begin{center}
   \includegraphics[width=6.0in]{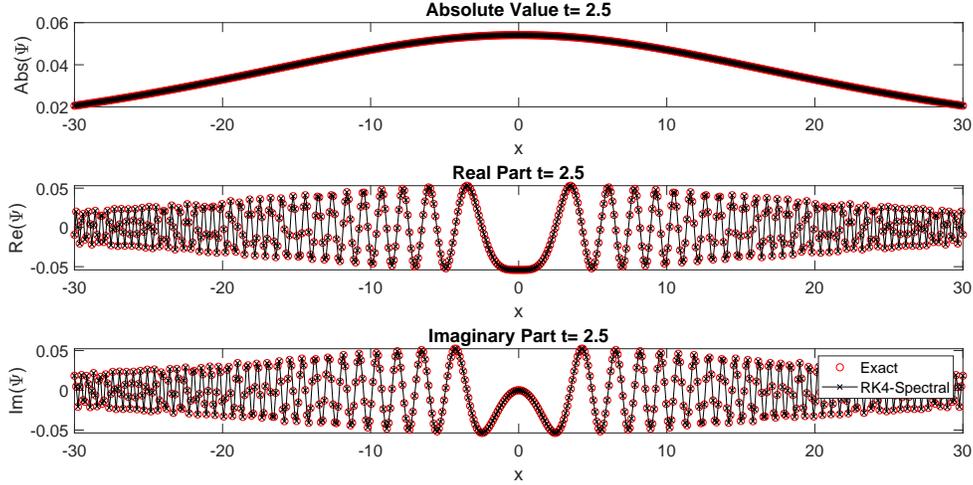}
  \end{center}
\caption{\small The waveform of the quantum chirp given by Eq.(\ref{eq01}) at the instant of $t=2.5$ for $A=1, \alpha_0=9.5, \phi_0=0$ a) absolute value b) real part c) imaginary part.}
  \label{fig1}
\end{figure*}

First of all, in order to test the validity and accuracy of the numerical scheme, we compare the analytical quantum chirp waveform given by Eqs.(\ref{eq06})-(\ref{eq08}) and against its numerical solution obtained by solving the NLSE given by Eq.(\ref{eq01}), in Fig.~\ref{fig1}. The subplots a, b and c in Fig.~\ref{fig1} show the absolute value, the real part and the complex part of the quantum chirps's wavefunction, respectively. As mentioned above, the numerical solution is obtained using the spectral scheme with a $4^{th}$ order Runge-Kutta time integrator. The simulation parameters are selected as $A=1, \alpha_0=9.5, \phi_0=0, dt=10^{-5}$ as mentioned before. The comparison depicted in Fig.~\ref{fig1} is obtained for the free evolution of the quantum chirp, in which no Zeno observations takes place. The initial non-dimensional time for these solutions is selected as $t=2.00$ and the time stepping is performed until the time of $t=2.50$. The actual domain of computations is selected to be $x \in [-60,60]$ as mentioned above, however only the part $x \in [-30,30]$ is depicted in  Fig.~\ref{fig1} for better visualization purposes. As Fig.~\ref{fig1} confirms, the numerical solution remains bounded and stable. We have also tested the numerical scheme and observed the same convergent behavior for much longer time scales considered.

\begin{figure*}[htb!]
\begin{center}
   \includegraphics[width=6.0in]{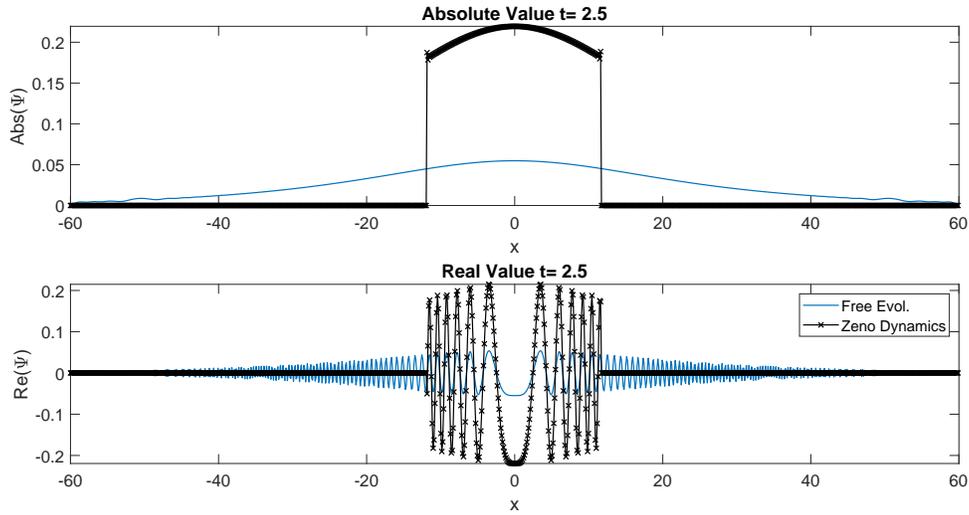}
  \end{center}
\caption{\small Comparison of the free evolution dynamics and the Zeno dynamics of the quantum chirp given by Eq.(\ref{eq01}) for $A=1, \alpha_0=9.5, \phi_0=0$ a) absolute value b) real part.}
  \label{fig2}
\end{figure*}

In Fig.~\ref{fig2}, we depict the waveform of the quantum chirp formulated by Eqs.(\ref{eq06})-(\ref{eq08}) under the Zeno effect and compare it with the chirp waveform under free evolution. In Fig.~\ref{fig2}, the subplot a and b refers to the absolute value and the real part of the complex wavefunction, respectively. The computation parameters are as before. The Zeno observation domain is selected as $L=[-11.6,11.6]$, which corresponds to $200$ indices about the origin. In numerical simulations, the Zeno observations on the numerical solution is imposed by applying the cycle given in Eq.(\ref{eq09}). In this simulation the quantum chirp is subjected to positive Zeno measurements at evenly spaced times in the interval of $t=[2.00-2.50]$. Between successive positive measurements, the quantum chirp evolves freely in the computation domain where the evolution process is governed by the NLSE given in Eq.(\ref{eq01}). In order to achieve the correct probability distributions, the quantum chirp under zeno effect is normalized as mentioned in the cycle given in Eq.(\ref{eq09}).

\begin{figure*}[htb!]
\begin{center}
   \includegraphics[width=6.0in]{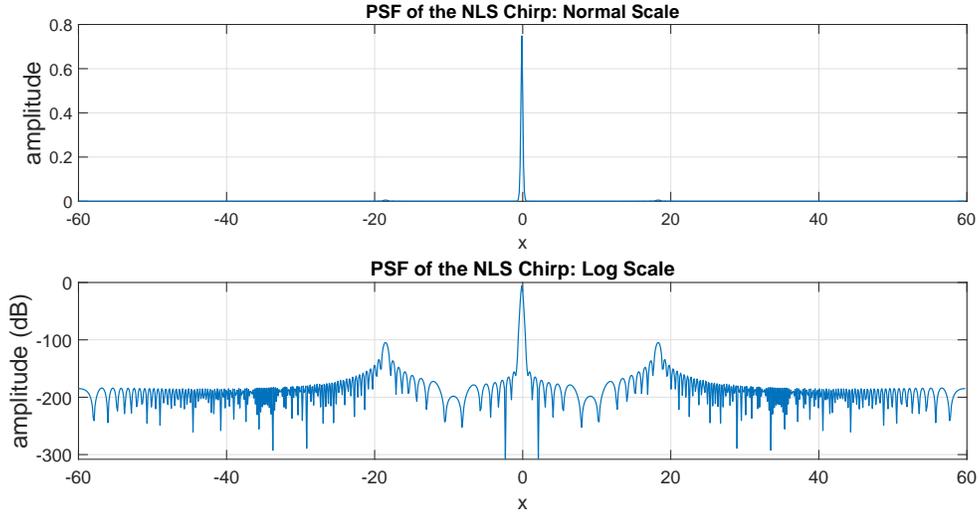}
  \end{center}
\caption{\small PSF of the freely evolving quantum chirp given by Eq.(\ref{eq01}) for $A=1, \alpha_0=9.5, \phi_0=0$ a) linear scale b) logarithmic scale.}
  \label{fig3}
\end{figure*}

\begin{figure*}[htb!]
\begin{center}
   \includegraphics[width=6.0in]{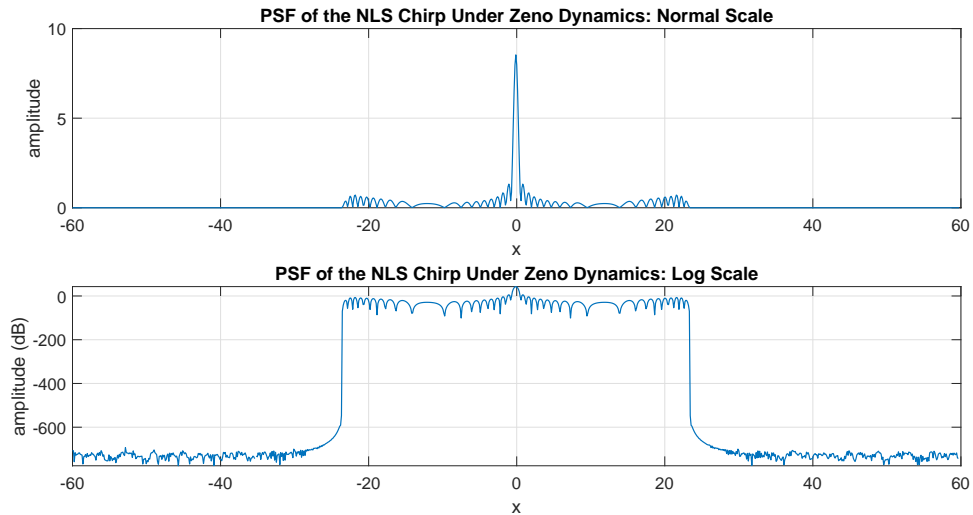}
  \end{center}
\caption{\small PSF of the quantum chirp given by Eq.(\ref{eq01}) for $A=1, \alpha_0=9.5, \phi_0=0$ under the effect of Zeno observations in the interval $[-11.6,11.6]$ a) linear scale b) logarithmic scale.}
  \label{fig4}
\end{figure*}

In our simulations, we observe that the evolution of the quantum chirp can be inhibited by the frequent Zeno measurements. Interestingly, we also observe that the quantum chirp profile preserves its chirped shape in the Zeno observation domain during its free evolution between two successive positive Zeno measurements, even for longer time scales than $t=2.50$. We also observe that, during the free evolution between two successive positive Zeno measurements, the quantum chirp exhibits a trapped behavior in the Zeno observation domain and the leakage or propagation of the chirp waveforms outward the observation domain is not observed. 

In order to analyze and assess the effects of Zeno observations of the signal and image resolution obtainable by the quantum chirp, we calculate its point spread functions (PSFs) for the cases of free evolution and inhibited evolution by Zeno dynamics. PSFs are commonly used to analyze imaging capabilities of various phase coded signals in optical communications, radar signal processing and similar branches. It is generally desired to have a PSF with a narrow mainlobe width and suppressed sidelobes to prevent ambiguity in imaging and target locations. Although for some chirps, the analytical PSF can be calculated exactly, after Zeno measurements the quantum chirp waveform is distorted by the measurement thus we calculate the PSFs numerically. Following \cite{Carrara, Collins, Curlander, Hein, Levanon, Soumekh1996, Soumekh2004, bay2013phd}, the PSF functions at any time of computation can be numerically calculated as
\begin{equation}
psf_{\psi}(x)=F^{-1} \left[|F[\psi]|^2 \right]=F^{-1} \left[|\widehat{\psi}(k)|^2 \right] 
\label{eq14}
\end{equation}
where $\widehat{\psi}(k)$ denotes the spectra of chirped waveform, $\psi(x,t)$, obtained by Fourier transforming it 
at a given instant. In Fig.~\ref{fig3} and Fig.~\ref{fig4}, we depict the PSF of the freely evolving quantum chirp and the PSF of the quantum chirp under the effect of Zeno observations. In both of these figures the first subplot is in normal and the second subplot is in logarithmic scale.

Comparing Fig.~\ref{fig3} and Fig.~\ref{fig4}, one can realize that the Zeno dynamics causes a significant distortion in the PSF, causing an increase in the -3dB mainlobe width, roughly from $0.23$ to $0.36$. Additionally, as seen in Fig.~\ref{fig4}, Zeno dynamics causes significant increase and widening of the sidelobes. Using these results, it is possible to conclude that the signal and image qualities attainable by quantum chirps can be significantly distorted by Zeno measurements. It is possible to jam the quantum chirps using these measurements with many possible applications in nonlinear quantum mechanics, optics, radar and quantum radar imaging. Additionally, the results presented in Fig.~\ref{fig3} and Fig.~\ref{fig4} are encouraging for a future analysis on the compression of quantum chirps using Zeno or anti-Zeno dynamics.

\begin{figure}[htb!]
\begin{center}
   \includegraphics[width=3.8in]{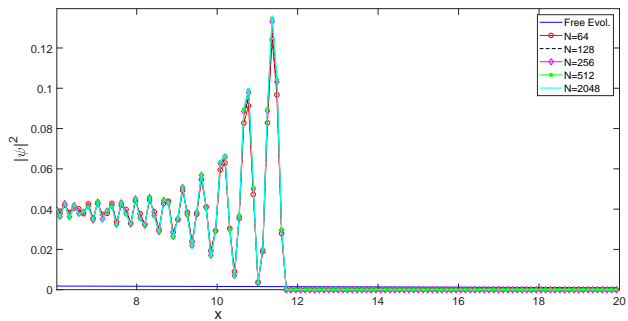}
  \end{center}
\caption{\small A comparison of the probability densities of the quantum chirp under free evolution and after $N=64, 128, 256, 512, 2048$ intermediate measurements taken in the observation domain of $[-11.6,11.6]$.}
  \label{fig5}
\end{figure}

In order to analyze and assess the effect of more frequent Zeno observations on the dynamics of the quantum chirp waveform, we depict the freely evolving wave profile and the wave profile with different Zeno measurement numbers, $N$, in Fig.~\ref{fig5}. For better illustration purposes, the results are plotted for the region $x \in [0,20]$, which is one sixth of the actual computation domain. As illustrated Fig.~\ref{fig5}, more frequent observations lead to slightly higher probability densities of lingering of the quantum chirp in the observation domain. The difference in the probability density levels are minor only. This is because of the fact that the quantum chirp under Zeno dynamics exhibits a trapped behavior in the observation domain and does not show a tendency for propagation or diffusion outward the observation domain independent of the number of Zeno measurements, $N$. Additionally, the chirped behavior is conserved in the domain of Zeno observations independent of the $N$. Outside the domain of Zeno observations,
the probability of lingering of the quantum chirp vanishes leading to a zero probability density.

\begin{figure}[h!]
\begin{center}
   \includegraphics[width=3.8in]{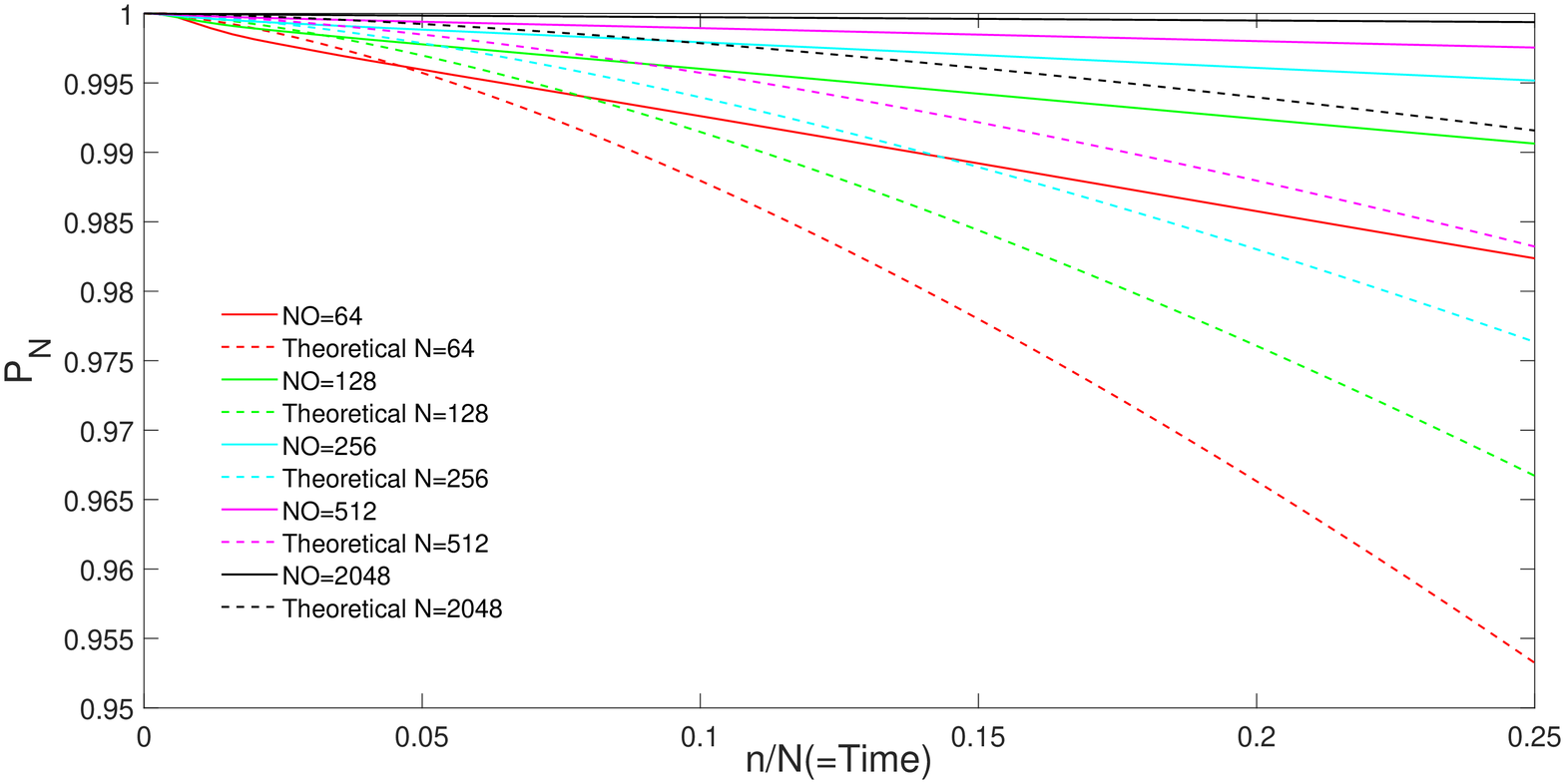}
  \end{center}
\caption{\small The cumulative probabilities of lingering of the quantum chirp in the observation domain $[-11.6,11.6]$ after $N$ intermediate measurements at the time of $t=n/N=0.25$.}
  \label{fig6}
\end{figure}

In Fig.~\ref{fig6}, we depict the cumulative probabilities of lingering of the quantum chirp in the observation domain evolved for a time interval of $t=[2.00, 2.50]$ and exposed to Zeno measurements in the time interval of $t=[2.25, 2.50]$, where this temporal interval differs to allow for the adjustment of the numerical scheme. The dashed lines in Fig.~\ref{fig6} denote the analytical cumulative distributions defined by the Eqs.(\ref{eq10})-(\ref{eq11}), whereas the continuous lines denote numerical findings. Although it is possible to observe that the cumulative probabilities of lingering of the quantum chirp increases as number of Zeno measurements increase both analytically and numerically, it is possible to realize that deviations of the analytical cumulative probability distributions from their numerical counterparts are not small as depicted in Fig.~\ref{fig6}. One of the reasons for these discrepancies is the fact that the analytical calculations rely on the optical wave analogy of the linear quantum mechanics and they are obtained using the linear Schr\"{o}dinger equation. However, the numerical simulations rely on the quantum chirp solution of the NLSE given by Eqs.(\ref{eq01}). The nonlinearity in the governing equation, the gain term, as well as the trapped behavior of the quantum chirp in the domain of Zeno observations significantly affect the cumulative probabilities depicted in Fig.~\ref{fig6}. Additionally, it is useful to note that although it is relatively smaller, the domain of Zeno observations and its length can affect these probabilities depending on the chirp waveform and its compression characteristics.

\section{Conclusion}

In this paper, we have analyzed the Zeno dynamics of quantum chirps in the frame of the NLSE with a gain/loss term. In particular, we have analyzed the Zeno dynamics of a chirped soliton solution of the NLSE having an envelope in the form of a $sech$ function. We have numerically solved the NLSE using a spectral scheme with a $4^{th}$ order Runge-Kutta time integrator and showed that evolution of the quantum chirps can be inhibited by using Zeno observations. We have showed that in the domain of Zeno observations, the chirp behavior and chirp waveform shape is preserved for long times scales. Additionally, in our numerical simulations we have observed that quantum chirps exhibit a trapped behavior in the observation domain during the temporal evolution between two successive Zeno measurements. Due to Zeno measurements
the point spread functions are significantly affected, causing an increase in the -3dB mainlobe width and an increase in the sidelobe levels which eventually leads to a reduction in the image resolution attainable by the quantum chirp. We have also investigated the lingering probabilities of quantum chirps after successive positive Zeno measurements. We have showed that nonlinearity of the governing equation, its gain/loss term and the trapped behavior of the quantum chirp in the Zeno observation domain significantly affect and increase the lingering probabilities of the quantum chirps in the observation domain, which lead to deviations from the analytical results obtained for the same lingering probabilities derived using the theory of linear quantum mechanics and optical wave analogy. 

Our findings and approach can be used to analyze the Zeno dynamics of quantum chirps in nonlinear quantum mechanics and nonlinear optics. The immediate applications of our findings is the jamming the quantum radar chirps and distorting the compression characteristics of optical pulses by Zeno dynamics. Our findings may lead to a different approach for chirp compression which depends on Zeno and anti-Zeno dynamics. Additionally, the dispersion and breaking characteristics of the quantum chirps under the effect of various types of optical potentials can be manipulated and enhanced using these dynamics. In our future work, we plan to investigate the revival characteristics of the quantum chirps, that is their characteristics after Zeno or anti-Zeno effects are ceased. One other possible research direction is to investigate these dynamics to impose time delays and/or phase singularities on the same or similar quantum chirps and investigate their characteristics. The procedure introduced in our paper can also be extended to model the Zeno dynamics of the many other fascinating nonlinear phenomena observed in the frame of NLSE and some of its extensions, which will definitely be helpful to advance the fields of nonlinear quantum mechanics and optical science and technology.



\end{document}